\documentclass[twocolumn,prl,showpacs]{revtex4}

\usepackage{amsmath}
\usepackage{amsfonts}
\usepackage{amssymb}
\usepackage{epsfig}
\usepackage{dcolumn}
\usepackage{bm}

\begin{document}

\title{Holonomic quantum computation in decoherence-free subspaces}
\author{L.-A. Wu$^1$, P. Zanardi$^2$, and D.A. Lidar$^{1,3}$}
\affiliation{$^1$Chemistry Department and Center for
Quantum Information and Quantum Control, University of Toronto, 80 St.
George St., Toronto, Ontario M5S 3H6, Canada}
\affiliation{$^2$Institute for Scientific Interchange, Villa Gualino, Viale
Settimio Severo 65, I-10133 Torino, Italy}
\affiliation{Departments of Chemistry and Electrical Engineering,
  University of Southern California, Los Angeles, CA 90089}

\begin{abstract}
We show how to realize, by means of non-abelian quantum holonomies, a
set of universal quantum gates acting on
decoherence-free subspaces and subsystems. In
this manner we bring together the quantum coherence stabilization virtues of
decoherence-free subspaces and the fault-tolerance of all-geometric
holonomic control. We discuss the implementation of this scheme in the
context of quantum information processing using trapped ions and quantum dots.
\end{abstract}

\maketitle

\textit{Introduction}.--- The implementation of quantum information
processing (QIP) poses an unprecedented challenge to our capabilities of
controlling the dynamics of quantum systems. The challenge is
twofold and somewhat contradictory. On the one hand one must (i) maintain
as much as possible the isolation of the computing degrees of freedom from the
environment, in order to preserve their \textquotedblleft
quantumness\textquotedblright ; on the other hand (ii) their dynamical
evolution must be enacted with extreme precision in order to avoid errors
whose propagation would quickly spoil the whole quantum computational
process. To cope with the decoherence problem (i), active strategies such as
quantum error correcting codes \cite{Steane:99}, as well as passive ones such as
error avoiding codes \cite{Duan:97PRLZanardi:97cLidar:PRL98}, have been contrived. The latter are based
on the symmetry structure of the system-environment interaction, which under
certain circumstances allows for the existence of decoherence-free subspaces
(DFS), i.e., subspaces of the system Hilbert state-space over which the
dynamics is still unitary. DFSs have been experimentally
demonstrated in a host of physical systems (e.g.,
\cite{Kielpinski:01,Kwiat:00Mohseni:02Ollerenshaw:02Bourennanne04a}).
The DFS idea of symmetry-aided protection has been generalized to noiseless
subsystems \cite{Knill:99aZanardi:99dKempe:00}, experimentally tested
in Ref.~\cite{Viola:01b}.

Holonomic quantum computation
(HQC) \cite{ZanardiRasetti:99} is an all-geometric strategy
wherein QIP is realized by means of adiabatic non-abelian quantum holonomies 
\cite{Wilczek:84}. Quantum information is encoded in a degenerate eigenstate of a
Hamiltonian depending on a set of controllable parameters e.g., external
laser fields. When the latter are adiabatically driven along a suitable
closed path, the initial quantum state is transformed by a non-trivial
unitary transformation (holonomy) that is geometrical in
nature. Following the ion traps HQC-implementation proposal \cite{Duan-Science:01}, several
other schemes, based on a variety of physical setups, were proposed \cite{Recati:02Faoro:03Li:04Bernevig:05}. One expects the
geometrical nature of quantum holonomies to endow HQC with inherent robustness against
certain errors. This alleged fault-tolerance has only recently been
subjected to serious scrutiny \cite{Solinas:04Fuentes-Guridi:05}; the resulting, still
developing picture, is that while stability against decoherence must be
further assessed (indeed the adiabatic theorem was only recently generalized
to open quantum systems \cite{SarandyLidar:04}), HQC seems to exhibit a
strong robustness against stochastic errors in the control process
generating the required adiabatic loops \cite{Zhu:04}. From this point of view
HQC seems to be promising with respect to the general challenge (ii).

In this work we describe a QIP scheme which combines DFSs and HQC
\cite{HQC-DFS-note,carollo}. More specifically, we show how to perform
universal quantum computation 
within a two-qubit DFS for collective dephasing by using non-abelian
holonomies only. The discussion is then extended to consider general
collective decoherence as well. The appeal of such a strategy should, in
view of the above, be evident: try to bring together the best of two worlds,
namely the resilience of the DFS approach against environment-induced decoherence and
the operational robustness of HQC. Moreover, we formulate our results using
rather generic Hamiltonians, so that the scheme proposed in this work
appears to be a suitable candidate for experimental demonstration in a
variety of systems, including trapped ions and quantum dots.

\textit{Dark-states in a Decoherence-Free subspace}.--- Let us start by
considering a four-qubit system with state-space $\mathcal{H}_{4}\cong (\mathbf{C}^{2})^{\otimes \,4}$. We denote by $X_{l},Y_{l}$ and $Z_{l}$ the
three Pauli matrices acting on the $l$th qubit; for any pair $l\neq m$ of qubits we define the operators $R_{lm}^{x}:=\frac{1}{2}\left(
X_{l}X_{m}+Y_{l}Y_{m}\right) $, $R_{lm}^{y}:=\frac{1}{2}\left(
X_{l}Y_{m}-Y_{l}X_{m}\right) $, $R_{lm}^{z}:=\frac{1}{2}\left(
Z_{m}-Z_{l}\right) $. These operators have a non-trivial action over the
subspace of $\mathcal{H}_{lm}\cong \mathbf{C}_{l}^{2}\otimes \mathbf{C}_{m}^{2}$ spanned by $|0_l 1_m\rangle $ and $|1_l 0_m\rangle $
(where $|0\rangle/|1\rangle$ are the $+/-$ eigenstates of $Z$, respectively): they provide a
faithful representation of the su(2) algebra over this subspace, where
they act as the Pauli matrices, while they vanish on the orthogonal
complement spanned by $|0_l 0_m\rangle $ and $|1_l 1_m\rangle $. Let 
$Z:=\sum_{i=1}^{4}Z_{i}$, then $[R_{lm}^{\alpha
},Z\,]=0$ $\,(\forall l,m)$. It follows that every eigenspace of $Z$ is
invariant under the action of the $R_{lm}^{\alpha }$'s. In particular this
holds true for the subspace 
\begin{equation}
\mathcal{C}:=\mathrm{span}\{|1000\rangle ,|0100\rangle ,|0010\rangle
,|0001\rangle \}.  \label{calC}
\end{equation}
${\cal C}$ is a DFS against collective dephasing
\cite{Duan:97PRLZanardi:97cLidar:PRL98}, i.e., states in $\mathcal{C}$ are immune from decoherence induced
by system-bath interactions of the form $Z\otimes B$, where $B$ is an
arbitrary bath operator. Collective dephasing is
known to be a major source of decoherence in ion-trap based QIP \cite{Kielpinski:01}.

We assume that the system dynamics is generated by
\begin{equation}
H=\sum_{l>m}(J_{lm}^{x}R_{lm}^{x}+J_{lm}^{y}R_{lm}^{y}),  \label{eq1}
\end{equation}
where $l,m$ are qubit indices and the $J_{lm}$'s are \emph{controllable}
coupling constants. These are the parameters that will be driven along
controlled adiabatic loops to enact quantum gates via non-abelian holonomies.
When $J_{lm}^{y}=0$, $H$ is the XY Hamiltonian found in a variety of quantum
computing proposals, e.g., the quantum Hall proposal \cite{Mozyrsky:01},
quantum dots \cite{Imamoglu:99} and atoms in cavities
\cite{Zheng:00}. It also describes trapped ions subject to the
S{\o}rensen-M{\o}lmer scheme \cite{Sorensen:00}. The case $J_{lm}^{x}=0$
is 
related to the XY model via a unitary transformation.

By way of introduction to our HQC-DFS scheme, note that Hamiltonian (\ref{eq1}) has a hidden multi-level structure with interesting properties
(isomorphic to the one exploited in Ref.~\cite{Duan-Science:01}). Indeed, the Hamiltonian 
$H_{lmn}=\sum_{j=m,n}J_{jl}(R_{jl}^{x}\cos \varphi _{jl}+R_{jl}^{y}\sin
\varphi _{jl})$, in the basis $\{|e\rangle :=|100\rangle _{lmn},|g_{1}\rangle :=|010\rangle
_{lmn},|g_{2}\rangle :=|001\rangle _{lmn}\}$, with $l\neq m\neq n$, takes the form 
\begin{equation}
H_{lmn}=J_{lm}e^{i\varphi _{lm}}|e\rangle \langle g_{1}|+J_{ln}e^{i\varphi
_{ln}}|e\rangle \langle g_{2}|+\mathrm{{h.c}.}  \label{lmn}
\end{equation}
This is a so-called Lambda scheme, with $|e\rangle $ at the top and $|g_{1}\rangle ,|g_{2}\rangle $ at the bottom. It is well known that for
every value of the $J_{jl}$'s and $\varphi _{jl}$'s, Hamiltonian (\ref{lmn})
has one \emph{dark state} $|D(J_{lm},J_{ln})\rangle \propto J_{lm}e^{i\varphi _{lm}}|g_{2}\rangle
-J_{ln}e^{i\varphi _{ln}}|g_{1}\rangle $, i.e., a state
satisfying $H_{lmn}|D(J_{lm},J_{ln})\rangle =0$.

The key idea is now as follows: moving in the control parameter space to the
point $J_{lm}/J_{ln}=0$, one has that the dark state is given by $|g_{1}\rangle $; then, if one adiabatically changes the parameters along a
closed loop, at the end the state $|g_{1}\rangle $ will pick up a
non-trivial geometric phase. This simple fact, suitably generalized, is the
basic ingredient of our HQC scheme. We further supplement the dark state $|g_{1}\rangle $ by another state, 
that is also annihilated by the system
Hamiltonian (and thus does not acquire a dynamical phase either); these
states together will form our qubit. In other words, the crucial observation
is that one can embed within the DFS $\mathcal{C}$ [Eq.~(\ref{calC})] a manifold of dark states
that, for specific values of the coupling constants in Eq.~(\ref{eq1}),
coincide with logical encoded qubits. Holonomic manipulations are then used to generate a
universal set of quantum logic gates.

Let us stress that even if during state manipulation the system
described by Eq.~(\ref{eq1}) leaks
out of the logical encoding subspace 
(the states $|0\rangle _{L}$ and $|1\rangle _{L}$ defined below),
e.g., due to the breakdown of adiabaticity \cite{SarandyLidar:04}, this results just in the kind of errors that
HQC is robust against \cite{Solinas:04Fuentes-Guridi:05}. Moreover,
during such leakage the system \emph{never} abandons the DFS
$\mathcal{C}$. Thus protection against collective dephasing is
maintained throughout the 
whole gating period, which in turns allows one to stretch the gating time in such a way as to
fulfill the adiabatic constraint for a longer period of time than
would be possible without the DFS. 
These remarks at least partially counter a standard objection to HQC, that the use of slow gates gives decoherence more time to
exert its detrimental effects.

\textit{One-qubit holonomic gates}.--- We now show how to realize the
single-qubit phase gate $\exp (i\varphi Z_{L})$, where $Z_{L}|\alpha \rangle
=(-1)^{\alpha }|\alpha \rangle _{L}$ (note that $Z_{L}=R_{12}^{z}$). We
encode a logical qubit in $\mathcal{C}$ by $|1\rangle _{L}:=|0010\rangle $
and $|0\rangle _{L}:=|0001\rangle $, while the remaining pair $|a_{1}\rangle :=|1000\rangle $ and $|a_{2}\rangle :=|0100\rangle $ 
plays the role of ancillae. Let us set all the $J_{lm}$ to zero, except $J_{24}$ and $J_{34}$, such that the Hamiltonian (\ref{eq1})
reduces to 
\begin{equation}
H_{Z}=J_{24}(R_{24}^{x}\cos \varphi _{24}+R_{24}^{y}\sin \varphi
_{24})+J_{34}R_{34}^{x}.  \label{ham-Z}
\end{equation}
We can also write $H_{Z}=J_{34}|a_{1}\rangle \langle
a_{2}|+J_{24}e^{i\varphi }|a_{1}\rangle \langle 1|_{L}+\mathrm{h.c}$. This is a Lambda configuration with $|a_{1}\rangle $ at the top
and $|a_{2}\rangle ,|1\rangle _{L}$ at the bottom. Therefore, as in our
discussion above, $H_{Z}$ has a zero-eigenvalue eigenstate (dark state)
given by $|\Psi _{1}\rangle =\cos \theta |1\rangle _{L}-\sin \theta
e^{i\varphi }|a_{2}\rangle $ where $\theta =\tan ^{-1}(J_{24}/J_{34})$ and $\varphi =\varphi _{24}$. The state $|0\rangle _{L}$ is also a
zero-eigenvalue eigenstate that does not depend on the parameters $\theta$ and $\varphi$. 
By adiabatically changing $\varphi $ in such a way as to have a loop
starting from $\theta=\varphi =0$, the state $|1\rangle _{L}$ acquires a Berry phase
which is proportional to the solid angle $\Omega _{Z}$ swept out by the
vector $(\phi ,\theta )$ \cite{Berry:84}. Therefore after this
adiabatic loop one has that: $|0\rangle _{L}\rightarrow |0\rangle _{L}$ and $|1\rangle _{L}\rightarrow
e^{-i\Omega _{Z}/2}|1\rangle _{L}$, which is clearly equivalent to the
operation $\exp (i\Omega _{Z}Z_{L}/2)$. 

In order to obtain a universal set of gates we need to generate at least two
non-commuting single-qubit gates. Therefore we next show how to implement $\exp (i\varphi X_{L})=\exp (i\varphi R_{12}^{x})$. One way is to once again
establish an isomorphism between the system governed by Eq.~(\ref{eq1}),
restricted to $\mathcal{C}$, and the HQC model of Ref. \cite{Duan-Science:01}. Here we
provide an independent derivation. We turn on the couplings in such a way as
to obtain the Hamiltonian 
\begin{equation}
H_{X}=J_{34}R_{34}^{x}+J_{24}(\cos \varphi \frac{R_{24}^{x}-R_{14}^{x}}{\sqrt{2}}+\sin \varphi \frac{R_{24}^{y}-R_{14}^{y}}{\sqrt{2}}).
\label{ham-X}
\end{equation}
Let $|\pm \rangle _{L}:=(|1\rangle _{L}\pm |0\rangle _{L})/\sqrt{2}$. Then
one can readily check that under the action of $(R_{24}^{x}-R_{14}^{x})/\sqrt{2}$ the state $|+\rangle _{L}\mapsto 0$,$\,|-\rangle _{L}\mapsto
|a_{1}\rangle $, and $|a_{1}\rangle \mapsto |-\rangle _{L}$, etc. Therefore $H_{X}=J_{34}|a_{1}\rangle \langle a_{2}|+J_{24}e^{i\varphi }|a_{1}\rangle
\langle -|_{L}+\mathrm{h}$\textrm{.}$\mathrm{c}$. This is a Lambda
configuration with $|a_{1}\rangle $ at the top and $|a_{2}\rangle ,|-\rangle
_{L}$ at the bottom, so that $H_{X}$ supports a dark state $|\Psi
_{2}\rangle =\cos \theta |-\rangle _{L}-\sin \theta e^{i\varphi
}|a_{2}\rangle $ where $\theta =\tan ^{-1}(J_{24}/J_{34})$. The
similarity between $H_{X}$ and $H_Z$ is evident. Then, by executing an 
adiabatic loop in the parameter space in analogy to the $H_Z$ case, one obtains the geometric evolution $|+\rangle _{L}\rightarrow |+\rangle _{L}$, $|-\rangle _{L}\rightarrow
e^{-i\Omega _{X}/2}|-\rangle _{L}$, where now $\Omega _{X}$ is the solid angle
swept out by the vector $(\theta ,\varphi )$. Switching back to the
computational basis this transformation amounts to the map 
$|0\rangle _{L} \rightarrow \cos \frac{\Omega _{X}}{4}|0\rangle _{L}+i\sin 
\frac{\Omega _{X}}{4}|1\rangle _{L}$, and
$|1\rangle _{L} \rightarrow \cos \frac{\Omega _{X}}{4}|1\rangle _{L}+i\sin 
\frac{\Omega _{X}}{4}|0\rangle _{L}$,
which is equivalent to $\exp (i\Omega _{X}R_{12}^{x}/4)$.

\textit{Two-Qubit holonomic gates}.--- A crucial, and typically rather
demanding part of any QIP implementation proposal, is the realization of
an entangling two-qubit gate. We next consider this problem and demonstrate
how to solve it following a strategy relying on the same abstract holonomic
structure as that discussed above for one-qubit gates. The total state-space
is now $\mathcal{H}_{4}^{\otimes \,2}$, while the two-qubit code is spanned
by $|\alpha \rangle _{L}\otimes |\beta \rangle _{L}$,$\,(\alpha ,\beta
=0,1)$. The states $|a_{i}\rangle ^{\otimes \,2}\,$($i=1,2$) will
function as 
ancillae. Importantly, once again all the relevant states belong to a DFS
against collective dephasing.

Let us suppose that one can engineer the following controllable four-qubit
interaction 
\begin{eqnarray}
H_{4} &=&J_{24,68}(R_{24}^{x}\cos \varphi +R_{24}^{y}\sin \varphi
)(R_{68}^{x}\cos \varphi +R_{68}^{y}\sin \varphi )  \notag \\
&+&J_{34,78}R_{34}^{x}R_{78}^{x},  \label{ham-2qubit}
\end{eqnarray}
which should be recognizable as a straightforward extension of the one-qubit
Hamiltonian~(\ref{ham-Z}). Below we discuss the implementation of such a
Hamiltonian. To explicitly exhibit the dark state structure, we write this
Hamiltonian in the form 
$H_{4}=J_{34,78}R_{34}^{x}(|a_{1}\rangle \langle a_{2}|)^{\otimes
\,2}+J_{24,68}e^{i\varphi }(|a_{1}\rangle \langle 1_{L}|)^{\otimes \,2}+\mathrm{h.c.}$,
from which it is easily seen that there is one dark state, given by $\cos
\theta |1\rangle _{L}^{\otimes \,2}-\sin \theta e^{i\varphi }|a_{2}\rangle
^{\otimes \,2}$, where $\theta =\tan ^{-1}(J_{24,68}/J_{34,78})$. After the
same kind of adiabatic cyclic evolution as described in the single-qubit
gate case, one obtains in $\mathcal{C}^{\otimes \,2}$ the controlled
phase-shift gate $CP=\mathrm{diag}(1,1,1,e^{i\Omega _{P}/2})$, where, as
usual, $\Omega _{P}$ is the solid angle swept out by the vector $(\theta
,\varphi )$.

\textit{Extensions and generalizations}.--- The scheme described so far for the case of collective dephasing is
straightforwardly scalable to an arbitrary number $N$ of encoded DF qubits.
The total space is now given by $\mathcal{C}^{\otimes \,N}\cong (\mathbf{C}^{4})^{\otimes \,N}$, with $\mathcal{C}$ given in Eq.~(\ref{calC}). This, of
course, is still an eigenspace of the collective spin $z$-component, i.e., $Z=\sum_{l=1}^{4N}Z_{l}$. Following the procedure established above, the
controllable Hamiltonian used to generate a controlled phase-shift between
the $i$-th and the $j$-th encoded qubits, has the same structure as $H_{4}$
[Eq.~(\ref{ham-2qubit})], where, with obvious notation, $J_{24;68}\rightarrow J_{4(i-1)+2,4(i-1)+4;4(j-1)+2,4(j-1)+4}$ and $J_{34;78}\rightarrow J_{4(i-1)+3,4(i-1)+4;4(j-1)+3,4(j-1)+4}$, and similarly
for the $R_{kl}^{x},R_{kl}^{y}$ operators.

Next, by making use of 
\emph{noiseless subsystems} (NS) \cite{Knill:99aZanardi:99dKempe:00},
we show that our combined HQC-DFS strategy can also be applied
against \emph{general} collective decoherence \cite{Duan:97PRLZanardi:97cLidar:PRL98}. Our arguments will
be existential in 
nature, with constructive details to be discussed elsewhere. Let $\mathcal{H}_{J}\cong \mathbf{C}^{2J+1}$ denote the total spin-$J$ irreducible
representation of $su(2)$. The state-space of five qubits, i.e., $\mathcal{H}_{1/2}^{\otimes \,5},$ decomposes, with respect to the collective $su(2)$-representation (Clebsch-Gordan decomposition) as follows 
\begin{equation}
\mathcal{H}_{1/2}^{\otimes \,5}\cong \mathbf{C}^{4}\otimes \mathcal{H}_{3/2}\oplus \mathbf{C}^{5}\otimes \mathcal{H}_{1/2}\oplus \mathbf{C}\otimes 
\mathcal{H}_{5/2}.  \label{cl-go}
\end{equation}
Each of the $\mathbf{C}^{n_{J}}$ factors represents the multiplicity of the
total spin-$J$ irreducible representation \emph{and corresponds to a NS
against collective general decoherence}
\cite{Knill:99aZanardi:99dKempe:00}. Consider the first
term in (\ref{cl-go}): the multiplicity factor $\mathbf{C}^{4}$ for the $J=3/2$ representation provides a four-dimensional NS. It might then encode
two noiseless qubits, but, since we wish to perform QIP with holonomies, we
will instead use this $\mathbf{C}^{4}$ space as a code for just one
noiseless qubit $|\tilde{\alpha} \rangle _{L}$, ($\alpha =0,1$) and two
ancillary states $|\tilde{a}_{i}\rangle $, ($i=1,2$). Suppose now that
one is able to enact the controllable Hamiltonian $H_{\mathrm{NS}}=J^{\prime
}|\tilde{a}_{1}\rangle \langle \tilde{a}_{2}|+J_{0}^{\prime \prime }e^{i\varphi
}|\tilde{a}_{1}\rangle \langle \tilde{0}|_{L}+J_{1}^{\prime \prime }e^{i\varphi
}|\tilde{a}_{1}\rangle \langle \tilde{1}|_{L}+\mathrm{h.c.}$, which, when $J_{0}^{\prime \prime }=0$, admits a dark state given
by $|\Psi _{3}\rangle =\cos \theta |\tilde{1}\rangle _{L}-\sin \theta e^{i\varphi
}|\tilde{a}_{2}\rangle $, where $\theta =\tan ^{-1}(J_{1}^{\prime
  \prime }/J^{\prime})$. By resorting to the same considerations as previously developed for the
collective dephasing case, it should be clear that this allows us to enact a
phase gate $Z_{L}$ between $|\tilde{0}\rangle _{L}$ and $|\tilde{1}\rangle _{L}$. By
choosing $J_{0}^{\prime \prime },J_{1}^{\prime \prime }$ such that
$H_{\mathrm{NS}}$ becomes $J^{\prime }|\tilde{a}_{1}\rangle \langle
\tilde{a}_{2}|+J_{-}^{\prime \prime }e^{i\varphi }|\tilde{a}_{1}\rangle \langle \tilde{-}|_{L}
+\mathrm{h.c.}$ we can enact the $X_{L}$ gate, so that universal single-qubit
control by holonomies can be achieved in this case as well.

To realize the required multi-level controllable
Hamiltonian, we observe that the $\mathbf{C}^{4}$ space under
consideration is a four-dimensional irreducible representation of the
permutation group $\mathcal{S}_{5}$ [acting over the whole space as $\hat{\sigma}\otimes _{j=1}^{5}|\tilde{\alpha} _{j}\rangle =\otimes _{j=1}^{5}|\tilde{\alpha}
_{\sigma (j)}\rangle $, where$\,\sigma \in \mathcal{S}_{5}$]. Following Ref.~\cite{Zanardi:04}, universal control over this irreducible representation space
amounts to the ability to switch on and off a pair of \emph{generic}
Hamiltonians in the group algebra of $\mathcal{S}_{5}$. An important example
is provided by Heisenberg \emph{exchange} Hamiltonians, i.e., $\sum_{l<m}J_{lm}\mathbf{S}_{l}\cdot \mathbf{S}_{m}$ [where $\mathbf{S}_{l}=(X_{l},Y_{l},Z_{l})/2$], which are naturally available interactions in
several spin-based proposals for QIP \cite{Burkard:99,Kane:98Vrijen:00}.
The construction of two-encoded-qubit holonomic gates, and the
generalization to arbitrary numbers of such qubits, again follow the same
pattern as in the collective dephasing case.

\textit{Implementation}.---  We note that all required
Hamiltonians, $H_{Z}$, $H_{X}$ and $H_{4}$, have a similar form. These Hamiltonians all
involve control over both $\theta $ and $\varphi $. However, we are free to
choose any loop $C(\theta ,\varphi )$ in the $(\theta ,\varphi )$-parameter
space, and we can choose a loop which toggles between $\theta $ and
$\varphi$. For example, the loop $C(\theta ,\varphi ):$
$(0,0){\rightarrow }(\frac{\pi
}{2},{0}){\rightarrow }(\frac{\pi }{2},\varphi
_{0}){\rightarrow }(0,\varphi_0)\equiv(0,0)$ has 
this property, where $\varphi_0$ is the solid angle $\Omega$ swept out by the vector
$(\theta ,\varphi )$ for this specific loop.

Let us briefly address the feasibility of the control of $H_{Z}$, $H_{X}$
and $H_{4}$ in the context of actual quantum computing proposals. In
proposals based on electron spins in quantum dots \cite{Burkard:99}, the $H_{Z}$ and $H_{X}$ Hamiltonians are 
available when one takes into account
the spin-orbit interaction, for then the effective spin-spin interaction
becomes $H_{ij}(t)=J_{ij}(t)[\beta (t)(X_{i}Y_{j}-Y_{i}X_{j})+(1+\gamma(t))(X_{i}X_{j}+Y_{i}Y_{j})+Z_{i}Z_{j}]$
\cite{Kavokin:01,Stepanenko:04}. This Hamiltonian has enough degrees of freedom to
implement $H_{Z}$ and $H_{X}$, via separate control of $J$ and the dimensionless
anisotropy 
parameters $\mathbf{\beta }$ and $\gamma $, assuming the latter can be
made $\gg 1$, e.g., when the inter-dot distance is
large \cite{Kavokin:01}. The parameters 
$\mathbf{\beta }$ and $\gamma $ can be controlled via the confining 
potential or via pulse shaping, as has been discussed in
detail in Ref.~\cite{Stepanenko:04}, and $\varphi =\tan^{-1}\beta
/(1+\gamma )$. The parameter $\theta$ is controllable via, e.g., $J_{24}$ and $J_{34}$.
Implementation of $H_{4}$ is 
undoubtedly
more challenging, since four-body interactions are involved. Recent results
indicate that such terms arise by simultaneously coupling four quantum dots 
\cite{MizelLidar:04}; their relative strength can be controlled, e.g., by
adjusting the confining potentials \cite{WoodworthMizelLidar:05}.

In
proposals based on trapped ions, the implementation of $H_{Z}
$ and $H_{X}$ is directly possible using the S\o rensen-M\o lmer (SM) scheme 
\cite{Sorensen:00}. The SM gate between two ions implements $H_{Z}$
and $H_{X}$ with control over the various terms achieved via the phases of
two lasers \cite{Kielpinski:02LidarWu:02}. It is also possible to implement $H_{4}$ using the SM scheme, via control over two pairs of ions
\cite{Kielpinski:02LidarWu:02}. Given that a geometric two ion-qubit
phase gate has already been demonstrated \cite{Leibfried:03}, trapped ions seem to be particularly
favorable for the implementation of our proposed HQC-DFS scheme.

\textit{Conclusions}.--- We have combined the DFS and HQC techniques, and
showed how to realize universal quantum computation over a scalable DFS against
collective dephasing by using adiabatic holonomies only. We discussed an
extension to the general collective decoherence case, arguing that
controllability of exchange Hamiltonians would suffice. Remarkably, the
whole computational process is carried out \emph{within} the DFS. The
DFS embedding, along with the all-geometrical nature of HQC, promises
to give to this scheme a twofold resilience, against decoherence and stochastic control errors. The proposed universal
quantum gates are carried out via adiabatic manipulation of Hamiltonians
that are controllable in several proposals for QIP implementations. We are
therefore hopeful that the theoretical ideas presented here may stimulate
corresponding experimental activity.

\textit{Acknowledgements}.--- P.Z. acknowledges
financial support from the European Union FET project TOPQIP (Contract No.
IST-2001-39215) and discussions with A. Carollo. 
D.A.L. acknowledges financial
support from the DARPA-QuIST program and the Sloan Foundation.


\begin{thebibliography}{10}

\bibitem{Steane:99}
For a review see, e.g., {A.M. Steane},  in {\em {Introduction to Quantum Computation and Information}},
  edited by {H.K. Lo, S. Popescu and T.P. Spiller} ({World Scientific},
  Singapore, 1999), pp.\ {184--212}.

\bibitem{Duan:97PRLZanardi:97cLidar:PRL98}
{L.-M Duan and G.-C. Guo}, Phys. Rev. Lett. {\bf 79},  1953  (1997);
{P. Zanardi and M. Rasetti}, Phys. Rev. Lett. {\bf 79},  3306  (1997);
{D.A. Lidar, I.L. Chuang, and K.B. Whaley}, Phys. Rev. Lett. {\bf 81},  2594
  (1998).

\bibitem{Kielpinski:01}
{D. Kielpinski {\it et al}.}, Science {\bf 291},  1013  (2001).

\bibitem{Kwiat:00Mohseni:02Ollerenshaw:02Bourennanne04a}
{P.G. Kwiat {\it et al}.}, Science {\bf 290}, 498  (2000);
{M. Mohseni {\it et al}.}, Phys. Rev. Lett. {\bf 91},  187903  (2003);
{J.E. Ollerenshaw, D.A. Lidar, and L.E. Kay}, Phys. Rev. Lett. {\bf 91},
217904 (2003);
M. Bourennane {\it et al}.,  Phys. Rev. Lett. {\bf 92}, 107901 (2004).
  
\bibitem{Knill:99aZanardi:99dKempe:00}
{E. Knill, R. Laflamme, and L. Viola}, Phys. Rev. Lett. {\bf 84},  2525
(2000); P. Zanardi,  Phys. Rev.  A {\bf 63}, 012301 (2001); J. Kempe
{\it et al}., Phys. Rev. A {\bf 63}, 042307 (2001).

\bibitem{Viola:01b}
L. Viola {\it et al}., Science {\bf{293}}, 2059 (2001).

\bibitem{ZanardiRasetti:99}
  {P. Zanardi and M. Rasetti}, Phys. Lett. A {\bf 264},  94  (1999);
  {J. Pachos, P. Zanardi, and M. Rasetti}, Phys. Rev. A {\bf 61},  010305(R)
  (1999).

\bibitem{Wilczek:84}
{F. Wilczek and A. Zee}, Phys. Rev. Lett. {\bf 52},  2111  (1984).

\bibitem{Duan-Science:01}
{L.-M. Duan, J.I. Cirac, and P. Zoller}, Science {\bf 292},  1695  (2001).

\bibitem{Recati:02Faoro:03Li:04Bernevig:05}
See, e.g., {A. Recati {\it et al.}}, Phys. Rev. A {\bf 66},  032309  (2002);
{L. Faoro, J. Siewert, and R. Fazio}, Phys. Rev. Lett. {\bf 90},  028301
  (2003);
{Y. Li {\it et al.}}, Phys. Rev. A {\bf 70},  032330 (2004).
{B.A. Bernevig and S.-C. Zhang}, Phys. Rev. B {\bf 71},  035303 (2005).

\bibitem{Solinas:04Fuentes-Guridi:05}
{P. Solinas, P. Zanardi, and N. Zangh}, Phys. Rev. A {\bf 70},  042316 (2004);
{I. Fuentes-Guridi, F. Girelli, and E. Livine}, Phys. Rev. Lett. {\bf
  94}, 020503  (2005).

\bibitem{SarandyLidar:04}
{M.S. Sarandy and D.A. Lidar}, Phys. Rev. A {\bf 71},  012331  (2005).

\bibitem{Zhu:04}
{S.-L. Zhu and P. Zanardi}, eprint quant-ph/0407177.

\bibitem{HQC-DFS-note}
 An abstract connection between DFS and HQC was discussed in P. Zanardi,
  Phys. Rev. Lett. {\bf 87}, 077901 (2001). See also J.K. Pachos and A. Beige,
  Phys. Rev. A {\bf 69}, 033817 (2004), for a concrete implementation involving
  atoms trapped in resonant optical cavities.

\bibitem{carollo} Holonomies generated in DFSs dynamically following 
an adiabatically changing environment have been studied in A. Carollo
{\it et al.}, quant-ph/0507101; {\it ibid}, quant-ph/0507229.

\bibitem{Mozyrsky:01}
{D. Mozyrsky, V. Privman, and M.L. Glasser}, Phys. Rev. Lett. {\bf 86},  5112
  (2001).

\bibitem{Imamoglu:99}
{A. Imamoglu {\it et al}.}, Phys. Rev. Lett. {\bf 83},  4204  (1999).

\bibitem{Zheng:00}
{S.-B. Zheng and G.-C Guo}, Phys. Rev. Lett. {\bf 85},  2392  (2000).

\bibitem{Sorensen:00}
A. S$\o$rensen and K. M$\o$lmer, Phys. Rev. A {\bf 62},  022311  (2000).

\bibitem{Berry:84}
{M.V. Berry}, Proc. Roy. Soc. (Lond.) {\bf 392},  45  (1989).

\bibitem{Zanardi:04}
{P. Zanardi and S. Lloyd}, Phys. Rev. A {\bf 69},  022313  (2004).

\bibitem{Burkard:99}
{G. Burkard, D. Loss, and D.P. DiVincenzo}, Phys. Rev. B {\bf 59},  2070
  (1999).

\bibitem{Kane:98Vrijen:00}
{B.E. Kane}, Nature {\bf 393},  133  (1998);
{R. Vrijen {\it et al}.}, Phys. Rev. A {\bf 62},  012306  (2000).

\bibitem{Kavokin:01}
{K.V. Kavokin}, Phys. Rev. B {\bf 64},  075305  (2001);
{K.V. Kavokin}, Phys. Rev. B {\bf 69},  075302  (2004).

\bibitem{Stepanenko:04}
{D. Stepanenko {\it et al}.}, Phys. Rev. B {\bf 68}, 115306 (2003);
{D. Stepanenko and N.E. Bonesteel}, Phys. Rev. Lett. {\bf 93},  140501  (2004).

\bibitem{MizelLidar:04}
{A. Mizel and D.A. Lidar}, Phys. Rev. Lett. {\bf 92},  077903  (2004).

\bibitem{WoodworthMizelLidar:05}
{R. Woodworth, A. Mizel, and D.A. Lidar}, eprint quant-ph/0504165.

\bibitem{Kielpinski:02LidarWu:02}
{D. Kielpinski, C. Monroe, and D.J. Wineland}, Nature {\bf 417},  709  (2002);
{D.A. Lidar and L.-A Wu}, Phys. Rev. A {\bf 67},  032313  (2003).

\bibitem{Leibfried:03}
{D. Leibfried {\it et al.}}, Nature {\bf 422}, 415 (2003).

\end{thebibliography}
\end{document}